\documentclass[conference]{IEEEtran}
\ifCLASSINFOpdf
  \usepackage[pdftex]{graphicx}
  \graphicspath{{./Figuras/}{./Figuras/Plots/}{./Figuras}{./}}
  \DeclareGraphicsExtensions{.pdf,.jpeg,.png}
\else
  \usepackage[dvips]{graphicx}
  \graphicspath{{../eps/}}
  \DeclareGraphicsExtensions{.eps}
\fi
\usepackage{algorithmicx}
\usepackage{algpseudocode}
\usepackage[caption=false,font=footnotesize]{subfig}
\usepackage{url}

\usepackage{epstopdf}
\usepackage{tikz}
\usepackage{complexity}
\usepackage{array}
\usepackage{multirow}

\usepackage{pifont}

\usepackage[utf8]{inputenc}
\usepackage[T1]{fontenc}
\usepackage[english]{babel}
\usepackage{hyperref}

\begin{document}
%
\title{Pragmatic Requirements for Adaptive Systems: a Goal-Driven Modelling and Analysis Approach}

\author{\IEEEauthorblockN{Felipe Pontes Guimarães\IEEEauthorrefmark{1},
Genaina Nunes Rodrigues\IEEEauthorrefmark{2},
Raian Ali\IEEEauthorrefmark{3} and
Daniel Macêdo Batista\IEEEauthorrefmark{1}}
\IEEEauthorblockA{\IEEEauthorrefmark{1}University of São Paulo, Brasil\\Rua do Matão, 1010, São Paulo, SP -- Brasil\\\{felipepg,batista\}@ime.usp.br}
\IEEEauthorblockA{\IEEEauthorrefmark{2}Universidade de Brasília, Brasil\\Asa Norte, Brasília, DF CEP: 70910-900\\
genaina@cic.unb.br}
\IEEEauthorblockA{\IEEEauthorrefmark{3}Bournemouth University, UK \\Fern Barrow BH12 5BB\\rali@bournemouth.ac.uk}}


\maketitle

\begin{abstract}
Goal-models (GM) have been used in adaptive systems engineering for their ability to capture the different ways to fulfill the requirements. Contextual GM (CGM) extend these models with the notion of context and context-dependent applicability of goals. In this paper, we observe that the interpretation of a goal achievement is itself context-dependent. Thus, we introduce the notion of Pragmatic Goals which have a dynamic satisfaction criteria. We also developed and evaluated an algorithm to decide the Pragmatic CGM's achievability. Finally, we performed several experiments to evaluate and to compare our algorithm against human judgment and concluded that the specification of context-dependent goals' applicability and interpretations make it hard for domain stakeholders to decide whether the model covers all possibilities, both in terms of time and accuracy, thus showing the importance and contribution of our algorithm.

\end{abstract}


%
\IEEEpeerreviewmaketitle

\section{Introduction}

Goal-Models (GM) are well established requirements engineering tools to depict and break-down systems using socio-technical concepts like inter-dependent actors, goals, quality goals or softgoals, tasks and resources \cite{yu1998goals}. GM facilitate the understanding of the system as a whole and provide a rationale about why the system needs to execute certain functionality and their possible variations~\cite{yu1998goals}. In other words, it provides the goals for which the system should be designed and the various possible ways to reach those goals. 

The variability of goal achievement strategies is the baseline for an actor to adapt by deciding which alternative to adopt as a response to certain triggers or adaptation drivers, e.g. faults, errors, availability of computational resources and newly available services and packages. The dynamic environment in which the system operates, \textit{i.e.} its context, could also be an adaptation driver. The Contextual Goal Model (CGM) \cite{Raian2010} extends the traditional goal model~\cite{tropos,castro2002towards} with the notion of context. Context may be an activator of goals, a precondition on the applicability of certain alternatives to reach a goal and a factor to consider when evaluating the quality provided by each of these alternatives. 

However, we advocate another effect of context on CGMs and requirements in general. The interpretation of a goal achievement is itself context dependent. This means that, in certain contexts, the mere achievement of the sub-goals in a goal model does not imply that the parent goal has been achieved. As an example, consider an ambulance dispatch. The goal of arriving at the patient's location in timely fashion would be seen as achieved when this takes 15 minutes and he/she suffers from dizziness. However, the same goal would not be achieved if the patient suffered from a heart condition. The pragmatism, i.e. dynamic interpretation, is not about the quality but the boolean decision whether a goal is achieved. 

In this paper, we introduce the concept of \textit{pragmatic goals} to grasp and model the idea that a goal's interpretation varies according to context. We define the achievability of pragmatic goals as being the capability of fulfilling a goal as interpreted within the context of operation. We also develop and implement an algorithm to compute the execution plan which is likely to achieve a pragmatic goal in a certain context. 

We evaluate the applicability of our modeling and the necessity for a reasoning algorithm by applying it on a case study of a Mobile Personal Emergency Response System and comparing the performance and reliability of the answers generated by human volunteers and by our algorithm. Results showed that volunteers took up to 17 minutes to provide answers with 27.81\% reliability whereas our algorithm provided correct answers in just a few milliseconds. Finally, we performed a scalability analysis to show the usability of our algorithm in pinpointing context sets in which the CGM as a whole may become unachievable, as well as the possibility of using it to support runtime adaptation by laying out an execution plan which is likely to achieve the necessary constraints.

The paper is organized as follows. Section~\ref{sec:cgm} presents the CGM concept on which our model is based. Section~\ref{sec:pragmatic} presents the pragmatic goals and pragmatic goal achievability concepts.  Section~\ref{sec:model} presents the proposed model and automated reasoning to decide the pragmatic achievability.  Section~\ref{sec:applicability} demonstrates the applicability of the modeling and analysis approach. Section~\ref{related} presents related work and  Section~\ref{sec:conclusion} concludes the paper and outlines our future work.


\section{The Contextual Goal-Model}
\label{sec:cgm}

Contextual Goal Model, proposed in \cite{Raian2010}, explicitly captures the relation between the goal model and dynamic environment of a system. It considers context as an adaptation driver when deciding the goals to activate and the alternatives - subgoal, task or delegation - to adopt and reach the activated goals. Context can also have an effect on the quality of those alternatives and this is captured through  the notion of contextual contribution to softgoals. In terms of syntax and modeling constructs, context can be correlated to certain variation points in the goal model. It is also analyzed through a technique called Context Analysis which allows to derive a formula, made of observable pieces of information (facts). This formula represents, in a measurable way, the condition whether a context holds.

Context is defined as the reification of the system's environment, 	\textit{i.e.}, the surrounding in which it operates \cite{finkelstein2001framework}. For goal models, context is defined as a partial state of the world relevant to an actor's goals \cite{Raian2010}. An actor is an entity that has goals and can decide autonomously how to achieve them. A context may be the time of a day, a weather condition, patient's chronic cardiac problem, etc.

\begin{figure*}[hbtp]
	\begin{center}
		\includegraphics[width=.9\textwidth ]{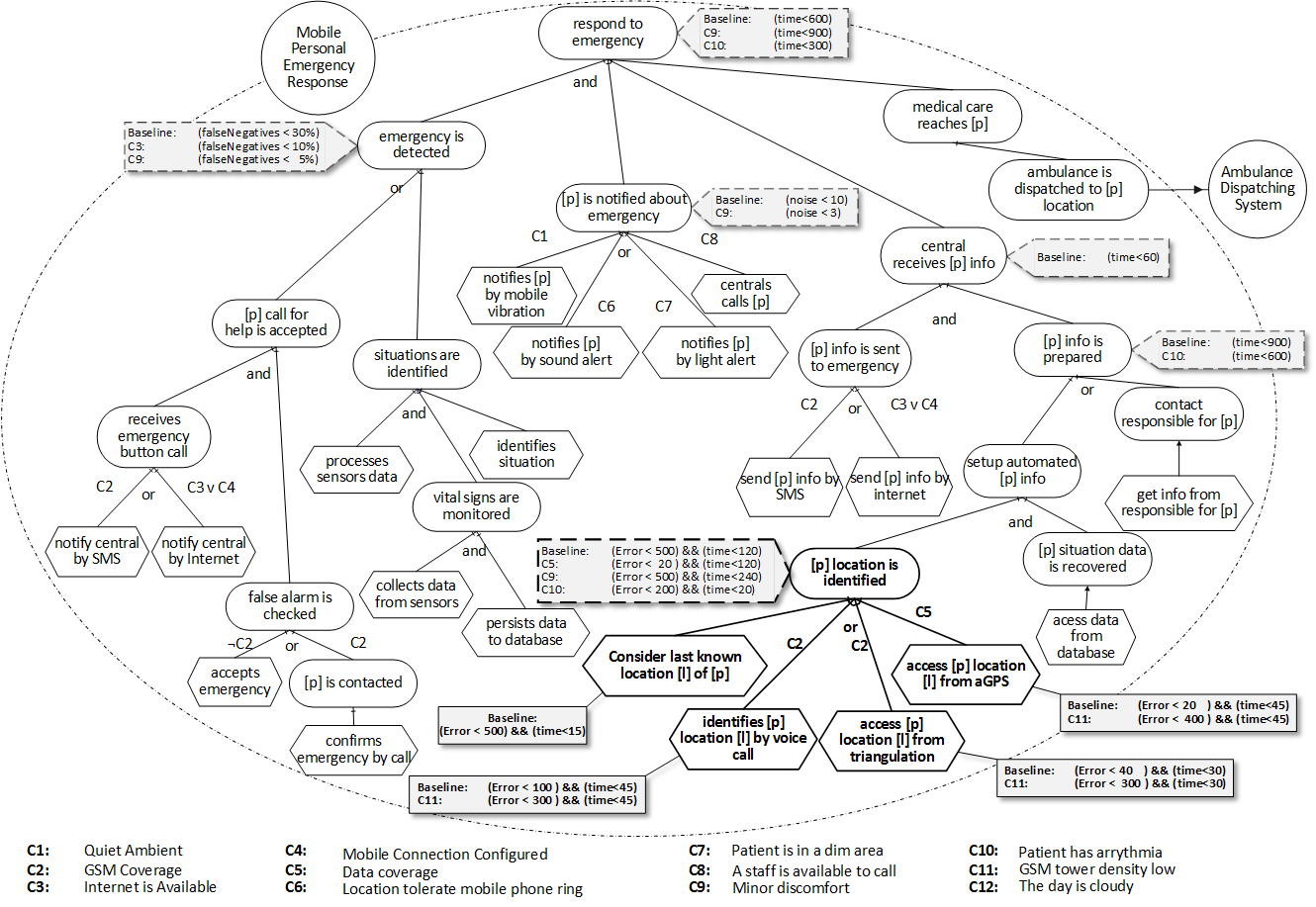}
		\caption{A CGM for responding to emergencies in an assisted living environment (adapted from \cite{Mendonca2014})}
		\label{fig:cgm}
	\end{center}
\end{figure*}

The CGM presented in Figure \ref{fig:cgm} depicts the goals to be achieved by a Mobile Personal Emergency Response System which is meant to respond to emergencies in an assisted living environment. The root goal is ``respond to emergency", which is performed by the actor \texttt{Mobile Personal Emergency Response}. The root goal is divided into 4 subgoals: ``emergency is detected", ``[p] is notified about emergency", ``central receives [p] info" and ``medical care reaches [p]'' ([p] stands for ``patient''). Such goals are then further decomposed, within the boundary of an actor, to finally reach executable tasks or delegations to other actors. A task is a process performed by the actor and a delegation is the act of passing a goal on to another actor that can perform it.

For instance, the goal ``setup automated [p] info" is decomposed into two subgoals: ``[p] location is identified" and ``[p] situation data is recovered". Such subgoal may then be realized via the task ``access data from database". An example of delegation can be seen at the goal ``ambulance is dispatched to [p] location". This goal is not performed by the \texttt{Mobile Personal Emergency Response} system but rather delegated to another actor which is the \texttt{Ambulance Dispatching System}.

The CGM observes that not all the subgoals, delegations and/or tasks are always applicable. Some of them depend on certain contexts whether they hold. For instance, the task ``identifies [p] location [l] by voice call" is only applicable if there is GSM (Global System for Mobile Communications) coverage at [p]'s location. The part of the CGM in which only applicable refinements, or nodes, remain is called the CGM variant for the current context. The possible contexts (C1-C12) are listed in the lower side of Figure~\ref{fig:cgm}.

\section{Pragmatic Requirements}
\label{sec:pragmatic}

Traditionally in a CGM, achieving one (OR-Decomposition) or all (AND-Decomposition) of the subgoals is seen as a satisfactory precondition for achieving the parent goal. We argue that the achievement of some goals would need to be seen from a more pragmatic point-of-view and not as a straightforward implication of the achievement of other goals or the execution of certain tasks. The decision whether a goal is achieved could be context-dependent. Thus we need a more flexible definition of goals to accommodate their contextual interpretation and achievement measures.


The representation of the quality of achievement of a goal as a softgoal is different from the pragmatism of the goal achievement. The pragmatic nature of a goal is not a matter of achieving it with higher or lower quality, but achieving it at all. Also, it has to do with the context at the time of execution and not with the model itself, making the quality requirements more strict or even relaxing them when some contexts apply.

Take the example of Figure~\ref{fig:cgm}: in general, the ambulance may take up to ten minutes to arrive. However, for a patient with a minor discomfort it can take its time and arrive about ten minutes later without suffering any penalty. For this situation, there is no need to hurry since there is no life threat. On the other hand, if the ambulance takes the same ten minutes to reach a patient having a heart attack, one cannot say the goal was achieved. In these situations, the delivered level of quality may not be a separate part from the boolean answer of whether a goal is achieved or not. Such quality level is intricately integrated, in a quantifiable way,  into the very definition of the goal's achievement.


To be able to illustrate the pragmatic nature of some goals, let us consider the CGM from Figure~\ref{fig:cgm} and focus on the highlighted goal ``\texttt{[p]} location is identified'' (this goal will be called \textbf{G$_{loc}$}). The CGM states that there are up to four ways of achieving this goal: either by considering the last known location of the patient, by a voice call asking the patient on his/her location, by GSM signal triangulation, or by GPS lock. In the traditional goal model interpretation, given that at runtime the proper contexts hold, the execution of any  of these tasks is enough to consider \textbf{G$_{loc}$} achieved. In this conception, \textbf{G$_{loc}$} is a clear example of hard goal.

However, from a pragmatic point-of-view, the simple fact that the location was discovered may not be enough. A location estimate based on the GSM signal triangulation may have a very low precision, a voice call may be achieved but the patient may not know how to describe its position, and a GPS lock may take too long in some cases. These nuances could be accepted in certain contexts while in other contexts they may lead to consider the goal unsatisfied. As a result, the pragmatic requirements must come into play.


A pragmatic goal describes the means to achieve it and also the interpretation of that achievement. This interpretation, which depicts the goal's pragmatic fulfillment criteria, can be expressed as a set of mandatory and crisp, therefore quantifiable, quality constraints (QCs). Unlike softgoals, these QCs are quantified measures needed for the fulfillment of a goal and an inherent part of its definition. In the previous example, the goal \textbf{G$_{loc}$} could then be defined as: ``in order to reach goal \textbf{G$_{loc}$}, the location must be identified so within an error radius of maximum 500m and in less than 2 minutes''. Again, this would not suffice, as a radius of 500m and 2 minutes might be an over-relaxed condition for patients under critical conditions. On the other hand, setting the highest possible level of demand for all situations is likely unreal and could lead to a huge waste of resources.

This brings into light another aspect to be taken into account for the pragmatic requirements: the fact that the interpretation for the achievement of a goal is itself context-dependent. We consider that there is a basic default condition for achieving a goal. On top of that, and for specific contexts, we could relax or further strengthen condition which interprets whether a goal is achieved. We propose that the contextual QCs on the achievement of a goal should be captured together with the other effects of context in the CGM. One advantage of capturing the pragmatic goals within the CGM is to enable reasoning on the possibility of achieving a goal under the current context and current constraints. We differentiate these interpretations in the sense that a relaxation condition is not mandatory but a condition that further strengthen the QC must necessarily be considered. 

For instance, in the previous example, a QC of getting a location within 500m in less than 2 minutes is a default constraint. However, if the user has access to mobile data connection (context C5) then a much preciser location can be obtained from the GPS. Under these circumstances, a lock within 500m may seem like an over-relaxed constraint. For a patient with cardiac arrhythmia (context C10), a more strict QC is needed. Suppose that the system has to ensure that an ambulance reaches the patient's home within 5 minutes. Possibly, in this case, a faster but less precise location would be better suited. Such nuances in the interpretation of the goal are summarized in Table \ref{tab:qosReq}. In the three specific contexts, the interpretation must be different than the baseline. For example, the requirements for a minor discomfort (context C9) are more flexible than the requirements for an arrhythmia (C10). In this case, the interpretation of goal \textbf{G$_{loc}$} may be expressed by Table \ref{tab:qosReq} and also presented as a box near the goal ``[p] location is identified'' on Figure~\ref{fig:cgm}.

\begin{table}
	\caption{Quality requirements for successfully achieving \newline \texttt{[p] location is identified}}
	\begin{center}
		\begin{tabular}{l|ll}
			\textbf{Context}	& \multicolumn{2}{c}{\textbf{Interpretation}} \\\hline
			Baseline			& (error < 500m)&\&\& (time < 120s)\\
			C5 	& (error < 20m)   &\&\& (time < 120s) \\
			C9  & (error < 500m) &\&\& (time < 240s) \\
			C10 & (error < 200m)  &\&\& (time < \ 20s)\   \\
		\end{tabular}
	\end{center}
	\label{tab:qosReq}
\end{table}



To be able to differentiate traditional goals from goals where the delivered quality of service (QoS) defines the condition for the goal's achievement, we introduced pragmatic goals into the CGM. A pragmatic goal is a hard goal with an interpretation expressed as a set of context-dependent QCs, shown in the graph within the dashed boxes. Unlike quality attributes and softgoals, a QC describes a \textit{mandatory} condition for considering a goal achieved. Such QCs may be imposed by the goal itself, i.e. by definition, as well as by each individual actor to reflect its own interpretation of the goal's fulfillment in the different contexts. 

\subsection{Achievability of Pragmatic Goals}

Not only a goal's fulfillment interpretation may vary according to the context, the QoS levels delivered by the tasks, i.e. executable processes, may also themselves differ. The same task may provide different QoS when executed in different contexts. This is represented by the boxes linked to tasks\footnote{Due to space limitations only the four tasks under \textbf{G$_{loc}$} have their context dependent apparent. This does not mean they do not exist in the other tasks}. This will propagate and affect the overall provided QoS of the parent goals. 


The reasoning part of our algorithm (Subsection~\ref{subsec:algorithm}) considers that pragmatic goals can only be achieved if their provided QoS comply with the QCs specified for them where both are context-dependent. This means that we extend the basic  effect of context on a CGM to cover success and achievement criteria. Such expressiveness enables further analysis for a key  adaptation decision: how to reach our goals while respecting the QCs under the current context where the goals interpretation, the space of applicable alternatives to reach a goal and the QoS provided by the tasks are all context-dependent. 

This part also considers the situation where it may not  be possible to meet the QoS standards which meet the goal's interpretation through any of the applicable sub goals, tasks and/or delegations as they deliver not a static but a context-dependent QoS level. In such cases, we classify the goal as unachievable and the reasoning part can explain the reason. 

To explain our rationale, let us consider the goal \textbf{G$_{loc}$} and the contexts impacting on its interpretation. 
In the presented CGM, given any moment in which $\neg C5 \wedge \neg C9 \wedge \neg C10$ holds, the goal \texttt{[p] location is identified} is interpreted as met as long as the error margin is less than 500m precision and the location is given within 120 seconds. However, when $ C5 \wedge C9 \wedge C10$  holds, such quality is insufficient and the required quality standards to meet this goal are more strict. In this case, the error margin cannot be more than 20m and must be given within 20 seconds. Finally, when $\neg C2 \wedge \neg C5 \wedge C10$ holds, it would not be possible to achieve this goal because the only applicable refinement left would be considering last known location of \texttt{[p]}. This option, however, has a much higher error margin than the required 200 meters.

In the above example, the conclusion is that under a certain context the system may not be able to determine the patient's location with the required precision. This, in practice, does not mean doing nothing. The motivation to do this analysis is because having such knowledge beforehand would allow consideration of other strategies, like adding more alternatives to the same goal to cover a larger range of contexts. At runtime, this conclusion would lead to search for a better variant at a higher-level goal by choosing  another branch of an OR-decomposition, which is able to deliver the required quality standard. Therefore, our analysis is both meant for design-time - reasoning to evaluate and validate the comprehensiveness of the solution - and for runtime - searching for the right alternative to reach goals in a specific context.

\section{Pragmatic Goal Model}
\label{sec:model}

In this section, we concretize our extension to the CGM and elaborate on the new constructs we add as well as their semantics. We mainly enhance the CGM with context-dependent goal interpretations and the expected delivered QoS, which are also context-dependent, for tasks in order to reason about the achievability of the goals for which these tasks are executed. 

To model pragmatic goals and their interpretation, we have extended the CGM's concept of a goal. A goal is refined into subgoals, tasks and/or delegations which must be achieved or performed to meet the parent goal. We extended this concept so that a pragmatic goal can now have an interpretation in the form of QCs that have to be met in order to render the pragmatic goal achieved at any given context. Thus we provided a quantifiable measure for a goal which encompasses  the verifiable satisfaction criteria and their dynamics in the different contexts. 

\begin{figure}[htp]
	\centering
	\includegraphics[width=5.5cm, clip=true, trim= 5mm 5mm 5mm 5mm]{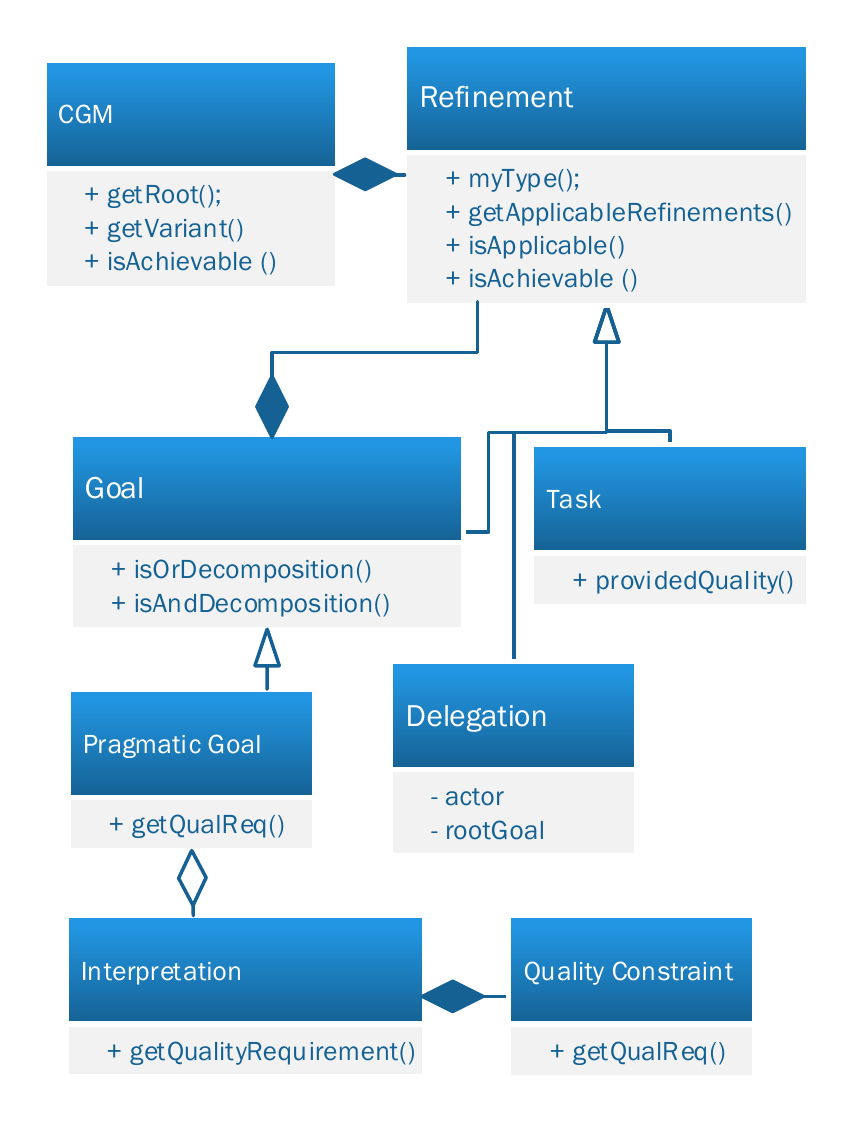}
	\caption{Conceptual metamodel for our extension to CGMs}
	\label{fig:uml}
\end{figure}

Figure~\ref{fig:uml} presents a  conceptual model of our extension to the CGM. For the focus of this paper, the CGM could be seen as an aggregation of \texttt{Refinements}. A \texttt{Refinement} may be specialized into several types: \texttt{Tasks}, \texttt{Delegations} and \texttt{Goals}. A \texttt{Delegation} represents when the \texttt{Goal} is pursued not by the current but by an external actor. \texttt{Tasks} are performed by the actor in order to achieve a  goal. \texttt{Tasks} may report the expected delivered quality for each metric through the \texttt{providedQuality} method. Goals have a \texttt{Refinements} set which define the subgoals, tasks and/or delegations that can be used for achieving it as well as a method to distinguish AND- from OR-compositions.

\texttt{Pragmatic Goals} extend the \texttt{Goal} concept with an \texttt{Interpretation}. A goal \texttt{Interpretation} is an abstract concept that has the function of cross-referencing a context and the appropriate \texttt{Quality Requirement} for that given context. The pragmatic goal is said to be achieved $iff$ such requirements are met. Otherwise, the goal's delivered QoS is considered inappropriate and the goal is not achieved regardless of achieving one or all of its \texttt{Refinements}. 

The \texttt{Quality Constraints} are expressed in terms of the \texttt{applicableContext} in which it holds, the \texttt{metric} that should be considered, the \texttt{threshold} which is a numerical value that represents the scalar value for such metric and the comparison which defines whether the threshold described is the maximum allowed value or the minimum. For instance, to state a quality requirement of at most 250ms for the execution time when context $C1$ holds, the \texttt{metric} would be ``ms", \texttt{threshold} would be 250, \texttt{condition} would be ``Less or Equal" and \texttt{applicableContext} would be $C1$.

Every Refinement inherits the \texttt{isAchievable} method. This method can be used either by the final users or by the higher level goals to define whether a particular goal can be achieved for a given quality requirement under the current context. Intermediate goals also have their own predefined Interpretation. While this is obviously necessary for the root goal, as the ultimate objective, we also allow certain subgoals to be defined as pragmatic. In principle, actors should be able to impose further constraints on the criteria for achieving any goal within their boundary. The importance of the subgoals quality requirement becomes obvious when dealing with delegation of goals where the external actor may have itself a different, more relaxed or more strict, quality constraint, not necessarily compatible with what the delegator intends.

Both the expectation of delivered quality by the tasks and the quality constraints for the goals, subgoals or delegations are added to the CGM. This is meant to be done by the requirements expert or the domain experts due to the need for specialized knowledge to define such metrics.  Different organizations may have different definitions of these interpretations and constraints which is yet another facet of the pragmatism of our approach.

\subsection{Achievability evaluation method }
\label{subsec:algorithm}
In this section we revisit the classical concept of achievability of a goal to fit the nature of Pragmatic CGMs. On top of the basic context effect on a CGM, we enable a higher model expressiveness. Such expressiveness will enable richer adaptation decisions which not only consider the static achievability but also the achievability under the dynamic context and its effect on the fulfillment criteria of a goal. The achievability of a goal and the space of adoptable alternatives to achieve it are essential information to plan adaptation, seen as a selection and enactment of a suitable alternative to reach a goal under a certain context. %

To evaluate the achievability of a particular pragmatic goal we present the algorithm in Figure~\ref{func:plan}. It implements the \texttt{Refinement} entity's \textit{isAchievable} method (Figure \ref{fig:uml}) and correlates three context-dependent aspects from the model:
(1) the applicable refinements; 
(2) the goals' interpretations and; 
(3) the delivered quality level provided by the tasks.

\begin{figure}[htbp]
	\begin{algorithmic}[1]
		\Require CGM, current context and desired QCs
		
		\State Goal root $\leftarrow$ cgm.getRoot()\label{algline:isRoot}
		\If{!root.isApplicable(current)\label{algline:isApplicableBegin}
		}
		\State \Return NULL\label{algline:returnNull} 
		\EndIf
		
		\If{(root.myType() == task)\label{algline:IfIsTask}}
		\If{(root.canFulfill(qualReq))\label{algline:compareQuality}
		}
		\State \Return new Plan(root)\label{algline:returnTask}
		\Else 
		\State \Return NULL \label{algline:returnTaskNull}
		\EndIf
		\EndIf
		
		\State QualityConstraint consideredQualReq
		\State consideredQualReq $\leftarrow$  
		\Statex root.interp.stricterQualityConstraint(root.qualReq, qualReq)\label{algline:stricterReq}
		
		\State Plan complete $\leftarrow$ NULL\label{algline:planDeclaration}
		\State deps $\leftarrow$ root.getRefinements(cgm, curContext)
		
		\ForAll{Refinement d \textnormal{\textbf{in}} deps \label{algline:forEachAndDecomp}
		}
		\State Plan p $\leftarrow$ d.isAchievable(cgm, context, \Statex consideredQualReq) \label{algline:isAchievableAnd}
		\If{(p != NULL)\label{algline:PlanIsNotNull}
		}
		
		\If{(root.isOrDecomposition())\label{algline:isOrDecomp}
		}
		\State \Return p \label{algline:returnPlanOr}
		\EndIf
		
		\If{(root.isAndDecomposition())\label{algline:isAndDecomp}}
		\State complete $\leftarrow$ addPlanToPlan(p, complete)\label{algline:addPlanAnd}
		\EndIf
		
		\ElsIf{(root.isAndDecomposition())\label{algline:planIsNull}
		}
		\State \Return NULL\label{algline:returnNullAnd}
		\EndIf
		
		\EndFor
		
		\State \Return{complete}\label{algline:isAndDecompEnd}
		
	\end{algorithmic}
	\caption{isAchievable(CGM cgm, Context current, QualityConstraint qualReq)}
	\label{func:plan} 
\end{figure}

The algorithm decides whether the root goal is achievable and, if so, lays out an execution plan, i.e.,  the set of all tasks to be executed, likely to achieve the desired QCs. The algorithm is recursive with a proven linear complexity with respect to the number of refinements in the CGM\footnote{Formal demonstration available at \url{https://github.com/felps/PragmaticGoals/blob/master/AlgorithmComplexity.pdf}. Accessed on January 22th, 2015}, building on the fact that the CGM is a tree-structured model without loops and that each refinement may be seen as a tree node.

The algorithm considers the root node of the CGM (line \ref{algline:isRoot}) and checks whether the root goal is itself applicable under the current context (line \ref{algline:isApplicableBegin}), returning NULL if it is not (line \ref{algline:returnNull}). In the particular case when the variant's root node is a task (line \ref{algline:IfIsTask}) it can readily decide on the achievability. This is because the task nodes know the expected QoS it can deliver for each metric under the context considered in the CGM. By comparing the delivered QoS and required QCs (line \ref{algline:compareQuality}), the node can decide whether it is capable or not of delivering such QCs. If it can, it will return a plan consisting only of this task (line \ref{algline:returnTask}), otherwise it will return \texttt{NULL} (line \ref{algline:returnTaskNull}) and indicate its inability to fulfill the goal's interpretation.

If the root is not a Task, the algorithm will define its quality requirement as the stricter \texttt{Quality Constraints} between its own and the QCs passed on as parameters (line \ref{algline:stricterReq}) and begin laying out an execution plan to fulfill such QCs (line \ref{algline:planDeclaration}).  For each of the applicable refinements, it will evaluate if it is achievable (line \ref{algline:isAchievableAnd}). If the refinement is achievable then, for OR-decompositions, the algorithm returns this plan immediately (lines \ref{algline:isOrDecomp} - \ref{algline:returnPlanOr}) and for AND-decompositions it is added to the \texttt{complete} plan (lines \ref{algline:isAndDecomp}-\ref{algline:addPlanAnd}). Otherwise, if the refinement is unachievable it will immediately return NULL for AND-decompositions (line \ref{algline:returnNullAnd}). Finally, for AND-decompositions, should all refinements are achievable it will return the \texttt{complete} plan (line \ref{algline:isAndDecompEnd}).

As an outcome, an execution plan is returned for achievable goals. For unachievable goals the NULL value is returned to indicate the inability of fulfilling the required constraints, allowing for alternate means of achieving higher level goals to be explored.

\section{Pragmatic Model and Achievability Algorithm Evaluation}
\label{sec:applicability}

In this Section, we aim to show the need for an algorithmic approach to handle Pragmatic Goals and to evaluate the capability of the proposed model to scale over the Pragmatic CGM size with regard to the amount of goals and contexts. 

To do so we used the Goal-Question-Metric (GQM) evaluation methodology~\cite{gqm}. GQM is a goal-oriented approach used throughout software engineering to evaluate products and software processes. It assumes that any data gathering must be based on an explicitly documented logical foundation which may be either a goal or an objective. 

GQM's first step is to define high-level goals to be evaluated. For each goal, a plan consisting of a series of quantifiable questions is devised to specify the necessary measures for duly assessing the evaluation [7]. These questions identify the necessary information to achieve the goals while the metrics define the operational data to be collected to answer each question.

In such a methodology, the main goals of our evaluation are: (I) to show the need for an algorithmic approach; (II) to evaluate the capability of using our approach for adaptive systems at runtime; (III) to evaluate whether our approach across may be used to identify context combinations which render the Pragmatic CGM's root goal unachievable by construction.

From these goals, the GQM plan was defined and is presented in Table \ref{tab:GQMPlan}. 

\begin{table*}[hbtp]
	\centering
	\caption{GQM devised plan}
	\label{tab:GQMPlan}
	\resizebox{\textwidth}{!}{
		\centering
		\begin{tabular}{|p{0.7\textwidth}|p{.25\textwidth}|}
			\hline
			\multicolumn{2}{|c|}{\textbf{Goal 1: Determine the necessity for an algorithmic approach}}\\\hline
			\multicolumn{1}{|c|}{\texttt{Question}}& \multicolumn{1}{|c|}{\texttt{Metric}}\\\hline
			
			1.1\hspace{0.4cm}\texttt{How long would a human take to come up with an answer?}
			& Time to produce an answer
			\\\hline
			
			1.2\hspace{0.4cm}\texttt{How reliable would an answer provided by a human be?} & Percentage of correct answers\\\hline
			
			1.3\hspace{0.4cm}\texttt{How long would the algorithm take to come up with an answer?}
			& Time to produce an answer
			\\\hline
			
			1.4\hspace{0.4cm}\texttt{How reliable would an answer provided by the algorithm be?} & Percentage of correct answers\\\hline
			
			\multicolumn{2}{l}{}
			\\\hline
			\multicolumn{2}{|c|}{\textbf{Goal 2: Evaluate the algorithm's runtime usage capability}}\\\hline
			\multicolumn{1}{|c|}{\texttt{Question}}& \multicolumn{1}{|c|}{\texttt{Metric}}\\\hline
			
			2.1\hspace{0.4cm}\texttt{How does the algorithm scale over the amount of goals in the model in average? }& Execution time\\\hline
			
			2.2\hspace{0.4cm}\texttt{How does the algorithm scale over the amount of contexts in the model in average?} & Execution time\\\hline
			
			2.3\hspace{0.4cm}\texttt{How does the algorithm scale over the amount of goals in the model in the worst case scenario? }& Execution time\\\hline
			
			2.4\hspace{0.4cm}\texttt{How does the algorithm scale over the amount of contexts in the model in the worst case scenario?} & Execution time\\\hline
			
			\multicolumn{2}{l}{}\\\hline
			\multicolumn{2}{|c|}{\textbf{Goal 3: Evaluate the algorithm's capability of pinpointing unachievable context sets}}\\\hline
			\multicolumn{1}{|c|}{\texttt{Question}}& \multicolumn{1}{|c|}{\texttt{Metric}}\\\hline
			
			3.1\hspace{0.4cm}\texttt{Can the algorithm cover all context sets for models with increasingly large models and reasonable context amounts? }& Context sets coverage\\\hline
			
			%
			%
			
			
		\end{tabular}}
	\end{table*}

	\subsection{Experiment Setup}
	
	The experiment setup consisted in two parts: the human capability for evaluating a Pragmatic model and the algorithm scalability analysis. These parts and their evaluations were engineered to provide the metrics demanded by the GQM plan (Table \ref{tab:GQMPlan})
	
	For the human capability evaluation, we performed an experiment with 55 volunteers from the fields of computer science, software, electronics and automotive engineering. The group would be given a 10 minutes explanation on the concepts concerning the Pragmatic CGM and a 5 minutes presentation of the CGM from Figure \ref{fig:cgm}, its 6 pragmatic goals (up to three different interpretations) and the provided QoS for all tasks (up to 3 context-dependent values). 
	
	After the explanation, the volunteers received 4 context sets (sets 1,3 and 4 were achievable while set 2 was not) and were asked to identify a set of tasks that could fulfill the CGM's pragmatic aspect under that context set whenever the goal is achievable or, otherwise, state the goal as unachievable. They were instructed to check the wall clock and write down the time once they came up with the solution.
	
	To limit the amount of necessary volunteer time, a limit of 25 minutes was set. After such deadline there would be no need to continue since we considered that taking more than 6 minutes to decide on a single context set for a rather small model would already be inadmissible and make it impossible for human judgment to be considered as a plausible alternative for finding unachievable scenarios.
	
	As for the algorithm evaluation, all experiments to evaluate the necessity, correctness and performance of the algorithm were implemented as automated tests under Java's JUnit framework\footnote{The evaluation mechanisms, the complete result set as well as the implementing code are available at \url{https://github.com/felps/PragmaticGoals}. Accessed on January 22th, 2015} . This guarantees that the evaluation is both effortless and repeatable. 
	
	Each time measurement was performed 100 times over each model and the amount of time was measured using Java's \texttt{System.nanoTime()} feature. This method provides an approximate measure of time with precision of at least 1 ms. The final measurement was the average of all 100 executions.
	
	
	The context set coverage evaluation was performed similarly, but instead of executing 100 executions, it would continuously execute until all the context sets were covered or until the elapsed time surpassed ten seconds.
	
	The algorithm from Figure~\ref{func:plan}  was implemented\footnote{Source code is available at \url{https://github.com/felps/PragmaticGoals}. Accessed on January 22th, 2015} using Java OpenJDK 1.7.0\_65 and the evaluation tests were performed on a Dell Inspiron 15r SE notebook equipped with a Intel Core i7 processor, 8GB RAM running Ubuntu 14.10, 64 bits and kernel 3.16.0-29-generic. We also used the \texttt{EclEmma} Eclipse's plugin for ensuring the tests' code coverage.
	
	\subsection{Goal 1: Determine the necessity for an algorithmic approach}
	
	We advocate that the CGM, though very useful in sharing the view and understanding of the problem among system developers and stakeholders, is just too complex to be merely evaluated by human judgment. On the other hand, it is very fast and very simple for a computer to perform a similar analysis and present reliable and comprehensive data.
	
	We conducted an experiment with volunteers from the University of Brasilia and compared to the algorithmic approach performance. 
	
	\subsubsection{\textbf{Questions 1.1 and 1.2 - The human perspective evaluation}}
	
	In order to answer questions 1.1 and 1.2, we performed an experiment with 55 volunteers in the fields of computer science, software, electronics and automotive engineering. The participants were familiar with the use of modeling, even if not necessarily within a software engineering environment but also in the automotive and product design. This added weight to our experiment as the practitioners of our approach are also those who are product designers. 
	
	The group had an explanation on the concepts concerning the Pragmatic CGM and a presentation on the CGM from Figure \ref{fig:cgm}, its 6 pragmatic goals and the provided QoS for all tasks (up to 3 context-dependent values). After that, any remaining doubts about the Pragmatic CGM concept and/or the CGM itself were clarified.
	
	After the explanation, the volunteers received 4 context sets (sets 1,3 and 4 were achievable while set 2 was not) and were asked to identify a set of tasks that could fulfill the CGM's pragmatic aspect under that context set whenever the goal is achievable or, otherwise, state the goal as unachievable.
	
	For this experiment, we measured the time the volunteers took to produce an answer for each set (question 1.1) and the correctness of the answers produced (question 1.2). 
	
	\paragraph{Question 1.1: How long does a human take to come up with an answer?}
	
\begin{figure}[hbtp]
		\centering
		\resizebox{.4\textwidth}{!}{
			\begin{tikzpicture}{0pt}{0pt}{514pt}{346pt}
			\clip(0pt,346pt) -- (386.946pt,346pt) -- (386.946pt,85.5269pt) -- (0pt,85.5269pt) -- (0pt,346pt);
			\begin{scope}
			\clip(51.1912pt,352.022pt) -- (384.687pt,352.022pt) -- (384.687pt,115.639pt) -- (51.1912pt,115.639pt) -- (51.1912pt,352.022pt);
			\color[rgb]{0,0,0}
			\draw[line width=2pt, line join=miter, line cap=rect](71.5172pt,308.359pt) -- (86.5734pt,308.359pt) -- (86.5734pt,173.606pt) -- (71.5172pt,173.606pt) -- (71.5172pt,308.359pt);
			\color[rgb]{0,0,0}
			\draw[line width=2pt, line join=miter, line cap=rect](77.5397pt,129.19pt) -- (80.5509pt,129.19pt);
			\draw[line width=2pt, line join=miter, line cap=rect](77.5397pt,319.652pt) -- (80.5509pt,319.652pt);
			\draw[line width=2pt, line join=miter, line cap=rect](79.0453pt,319.652pt) -- (79.0453pt,308.359pt);
			\draw[line width=2pt, line join=miter, line cap=rect](79.0453pt,129.19pt) -- (79.0453pt,173.606pt);
			\draw[line width=2pt, line join=miter, line cap=rect](71.5172pt,250.393pt) -- (86.5734pt,250.393pt);
			\draw[line width=2pt, line join=miter, line cap=rect](77.5397pt,129.19pt) -- (80.5509pt,129.19pt);
			\draw[line width=2pt, line join=miter, line cap=rect](77.5397pt,345.247pt) -- (80.5509pt,345.247pt);
			\draw[line width=2pt, line join=miter, line cap=rect](164.113pt,206.73pt) -- (179.169pt,206.73pt) -- (179.169pt,151.774pt) -- (164.113pt,151.774pt) -- (164.113pt,206.73pt);
			\draw[line width=2pt, line join=miter, line cap=rect](170.136pt,129.19pt) -- (173.147pt,129.19pt);
			\draw[line width=2pt, line join=miter, line cap=rect](170.136pt,245.876pt) -- (173.147pt,245.876pt);
			\draw[line width=2pt, line join=miter, line cap=rect](171.641pt,245.876pt) -- (171.641pt,206.73pt);
			\draw[line width=2pt, line join=miter, line cap=rect](171.641pt,129.19pt) -- (171.641pt,151.774pt);
			\draw[line width=2pt, line join=miter, line cap=rect](164.113pt,169.842pt) -- (179.169pt,169.842pt);
			\draw[line width=2pt, line join=miter, line cap=rect](170.136pt,122.415pt) -- (173.147pt,122.415pt);
			\draw[line width=2pt, line join=miter, line cap=rect](170.136pt,250.393pt) -- (173.147pt,250.393pt);
			\draw[line width=2pt, line join=miter, line cap=rect](256.709pt,183.393pt) -- (271.765pt,183.393pt) -- (271.765pt,151.774pt) -- (256.709pt,151.774pt) -- (256.709pt,183.393pt);
			\draw[line width=2pt, line join=miter, line cap=rect](262.732pt,142.741pt) -- (265.743pt,142.741pt);
			\draw[line width=2pt, line join=miter, line cap=rect](262.732pt,236.089pt) -- (265.743pt,236.089pt);
			\draw[line width=2pt, line join=miter, line cap=rect](264.237pt,236.089pt) -- (264.237pt,183.393pt);
			\draw[line width=2pt, line join=miter, line cap=rect](264.237pt,142.741pt) -- (264.237pt,151.774pt);
			\draw[line width=2pt, line join=miter, line cap=rect](256.709pt,176.617pt) -- (271.765pt,176.617pt);
			\draw[line width=2pt, line join=miter, line cap=rect](262.732pt,142.741pt) -- (265.743pt,142.741pt);
			\draw[line width=2pt, line join=miter, line cap=rect](262.732pt,250.393pt) -- (265.743pt,250.393pt);
			\draw[line width=2pt, line join=miter, line cap=rect](349.305pt,156.291pt) -- (364.361pt,156.291pt) -- (364.361pt,141.988pt) -- (349.305pt,141.988pt) -- (349.305pt,156.291pt);
			\draw[line width=2pt, line join=miter, line cap=rect](355.327pt,127.684pt) -- (358.339pt,127.684pt);
			\draw[line width=2pt, line join=miter, line cap=rect](355.327pt,199.202pt) -- (358.339pt,199.202pt);
			\draw[line width=2pt, line join=miter, line cap=rect](356.833pt,199.202pt) -- (356.833pt,156.291pt);
			\draw[line width=2pt, line join=miter, line cap=rect](356.833pt,127.684pt) -- (356.833pt,141.988pt);
			\draw[line width=2pt, line join=miter, line cap=rect](349.305pt,156.291pt) -- (364.361pt,156.291pt);
			\draw[line width=2pt, line join=miter, line cap=rect](355.327pt,125.426pt) -- (358.339pt,125.426pt);
			\draw[line width=2pt, line join=miter, line cap=rect](355.327pt,224.044pt) -- (358.339pt,224.044pt);
			\end{scope}
			\begin{scope}
			\color[rgb]{0,0,0}
			\pgftext[center, base, at={\pgfpoint{8.28094pt}{231.19pt}},rotate=90]{\fontsize{11}{0}\selectfont{\textbf{Time to produce an answer (Seconds)}}}
			\color[rgb]{0.192157,0.215686,0.223529}
			\pgftext[center, base, at={\pgfpoint{36.9878pt}{111.875pt}}]{\fontsize{11}{0}\selectfont{0}}
			\pgftext[center, base, at={\pgfpoint{31.1653pt}{157.044pt}}]{\fontsize{11}{0}\selectfont{200}}
			\pgftext[center, base, at={\pgfpoint{31.1653pt}{202.213pt}}]{\fontsize{11}{0}\selectfont{400}}
			\pgftext[center, base, at={\pgfpoint{31.1653pt}{246.629pt}}]{\fontsize{11}{0}\selectfont{600}}
			\pgftext[center, base, at={\pgfpoint{31.1653pt}{291.798pt}}]{\fontsize{11}{0}\selectfont{800}}
			\pgftext[center, base, at={\pgfpoint{26.5955pt}{336.966pt}}]{\fontsize{11}{0}\selectfont{1.000}}
			\color[rgb]{0,0,0}
			\draw[line width=1pt, line join=bevel, line cap=rect](51.1912pt,126.932pt) -- (47.4272pt,126.932pt);
			\draw[line width=1pt, line join=bevel, line cap=rect](51.1912pt,149.516pt) -- (47.4272pt,149.516pt);
			\draw[line width=1pt, line join=bevel, line cap=rect](51.1912pt,172.1pt) -- (47.4272pt,172.1pt);
			\draw[line width=1pt, line join=bevel, line cap=rect](51.1912pt,194.685pt) -- (47.4272pt,194.685pt);
			\draw[line width=1pt, line join=bevel, line cap=rect](51.1912pt,217.269pt) -- (47.4272pt,217.269pt);
			\draw[line width=1pt, line join=bevel, line cap=rect](51.1912pt,239.101pt) -- (47.4272pt,239.101pt);
			\draw[line width=1pt, line join=bevel, line cap=rect](51.1912pt,261.685pt) -- (47.4272pt,261.685pt);
			\draw[line width=1pt, line join=bevel, line cap=rect](51.1912pt,284.269pt) -- (47.4272pt,284.269pt);
			\draw[line width=1pt, line join=bevel, line cap=rect](51.1912pt,306.854pt) -- (47.4272pt,306.854pt);
			\draw[line width=1pt, line join=bevel, line cap=rect](51.1912pt,329.438pt) -- (47.4272pt,329.438pt);
			\draw[line width=1pt, line join=bevel, line cap=rect](51.1912pt,352.022pt) -- (47.4272pt,352.022pt);
			\draw[line width=1pt, line join=bevel, line cap=rect](51.1912pt,138.224pt) -- (47.4272pt,138.224pt);
			\draw[line width=1pt, line join=bevel, line cap=rect](51.1912pt,183.393pt) -- (47.4272pt,183.393pt);
			\draw[line width=1pt, line join=bevel, line cap=rect](51.1912pt,228.561pt) -- (47.4272pt,228.561pt);
			\draw[line width=1pt, line join=bevel, line cap=rect](51.1912pt,272.977pt) -- (47.4272pt,272.977pt);
			\draw[line width=1pt, line join=bevel, line cap=rect](51.1912pt,318.146pt) -- (47.4272pt,318.146pt);
			\draw[line width=1pt, line join=bevel, line cap=rect](51.1912pt,115.639pt) -- (44.4159pt,115.639pt);
			\draw[line width=1pt, line join=bevel, line cap=rect](51.1912pt,160.808pt) -- (44.4159pt,160.808pt);
			\draw[line width=1pt, line join=bevel, line cap=rect](51.1912pt,205.977pt) -- (44.4159pt,205.977pt);
			\draw[line width=1pt, line join=bevel, line cap=rect](51.1912pt,250.393pt) -- (44.4159pt,250.393pt);
			\draw[line width=1pt, line join=bevel, line cap=rect](51.1912pt,295.562pt) -- (44.4159pt,295.562pt);
			\draw[line width=1pt, line join=bevel, line cap=rect](51.1912pt,340.73pt) -- (44.4159pt,340.73pt);
			\draw[line width=1pt, line join=bevel, line cap=rect](51.1912pt,352.022pt) -- (51.1912pt,115.639pt);
			\pgftext[center, base, at={\pgfpoint{217.939pt}{86.2797pt}}]{\fontsize{11}{0}\selectfont{\textbf{Context Sets}}}
			\color[rgb]{0.192157,0.215686,0.223529}
			\pgftext[center, base, at={\pgfpoint{79.0394pt}{100.583pt}}]{\fontsize{11}{0}\selectfont{Set 1}}
			\pgftext[center, base, at={\pgfpoint{172.012pt}{100.583pt}}]{\fontsize{11}{0}\selectfont{Set 2}}
			\pgftext[center, base, at={\pgfpoint{264.608pt}{100.583pt}}]{\fontsize{11}{0}\selectfont{Set 3}}
			\pgftext[center, base, at={\pgfpoint{357.204pt}{100.583pt}}]{\fontsize{11}{0}\selectfont{Set 4}}
			\color[rgb]{0,0,0}
			\draw[line width=1pt, line join=bevel, line cap=rect](51.1912pt,115.639pt) -- (384.687pt,115.639pt);
			\pgftext[left, base, at={\pgfpoint{88.0791pt}{246.842pt}}]{\fontsize{11}{0}\selectfont{600}}
			\pgftext[left, base, at={\pgfpoint{182.933pt}{168.876pt}}]{\fontsize{11}{0}\selectfont{240}}
			\pgftext[left, base, at={\pgfpoint{274.024pt}{172.112pt}}]{\fontsize{11}{0}\selectfont{270}}
			\pgftext[left, base, at={\pgfpoint{367.373pt}{150.044pt}}]{\fontsize{11}{0}\selectfont{180}}
			\end{scope}
			\end{tikzpicture}
			
		}
		\caption{Average Time (in seconds) for volunteers to produce an answer to each context set}
		\label{fig:contextSets}
	\end{figure}
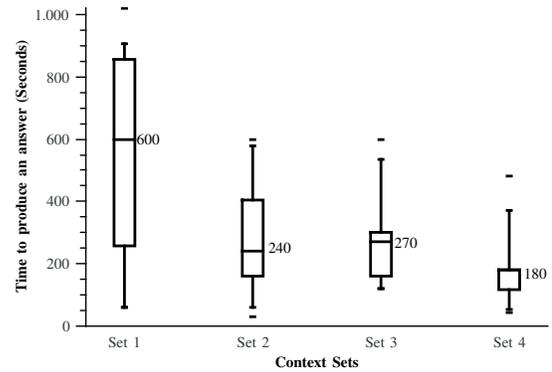

	The average time for a volunteer to produce a solution for each of the given context sets of the experiment is shown in Figure~\ref{fig:contextSets}. The box plots represent the median and the dispersion of the time (in seconds) it took the volunteers to come up with an answer for each provided set. These times varied a lot since it took a while for the participants to understand the whole idea and, as they progressed through the sets they gained experience thus becoming faster. Still in the best case (Context Set 4) most of them took between 100 and 180 seconds to come up with an answer.
	
	%
	
	\paragraph{Question 1.2: How reliable would an answer provided by a human be?}
	
	All the collected answers from the participants were then compared to the correct options. Out of the 97 answers produced, 54 were considered erroneous, 13 were partially correct and only 26 were precise. This means that 73.19\% of the provided answers were only partially correct or completely incorrect.
	
	Our experiment also showed that it was easier for the participants to correctly state a CGM as unachievable (75\% correct answers for set 3) than to find a task set which would satisfy the Pragmatic Requirements (11.6\% correct answers for sets 1, 2 and 4). This is probably because deeming a subgoal as unachievable may propagate this condition to the whole CGM or at least to an AND-decomposition sub-tree whereas finding a valid solution for the whole tree requires thoroughly investigating all options.
	
	\subsubsection{\textbf{Questions 1.3 and 1.4: The algorithmic approach evaluation}}
	
	The implemented algorithm receives three arguments as input: the CGM, the set of contexts and an optional user's quality constraint for the CGM's root goal. It outputs \texttt{NULL} if the root goal is unachievable or an execution plan, \textit{i.e.}, a set of tasks that can achieve the root goal. The execution plan must abide both by the quality constraint provided as input as well as any  pragmatic goals' interpretation. 
	
	\paragraph{Question 1.3: How long would the algorithm take to come up with an answer?}
	
	To evaluate the time for the algorithm execution on the CGM of Figure~\ref{fig:cgm}, we executed 1000 iterations of the algorithm for each context set. The results showed that the algorithm took, in average, less than 1 ms to be executed in each of the 4 scenarios.
	
	\paragraph{Question 1.4: How reliable would an answer provided by the algorithm be?}
	
	To validate the algorithm's correctness, we implemented tests for each context set used with the volunteers. For each one of these, we have identified all of the inapplicable tasks - both because of context or quality constraints - and asserted that the outputted execution plan did not contain any of these. All the test succeeded thus providing evidence of the algorithm correctness.

	\begin{figure}[hbtp]
		\centering
		\includegraphics[width=.48\textwidth]{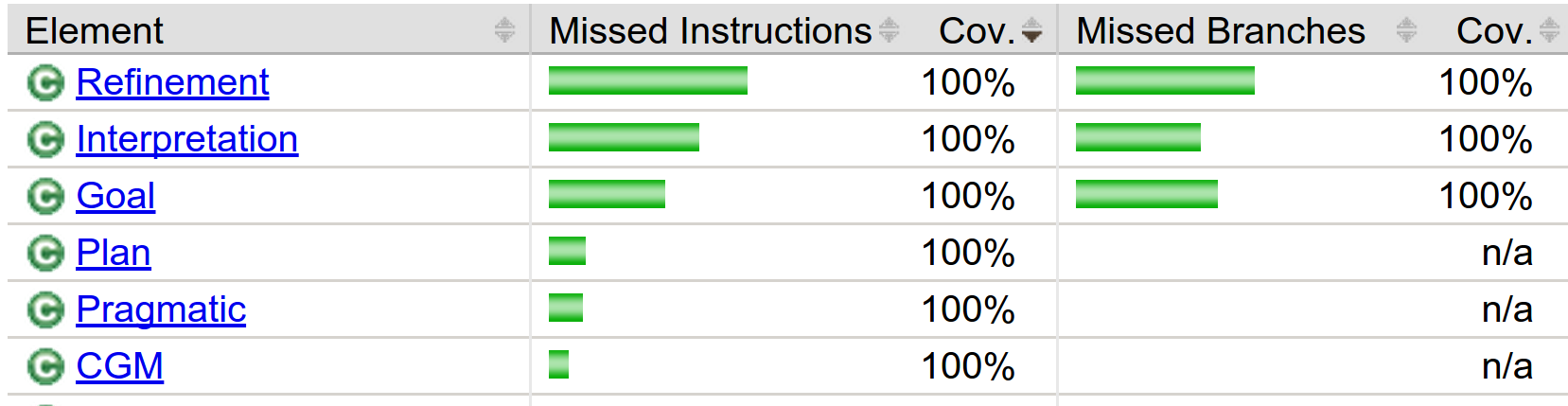}
		\caption{Eclipse's EclEmma plugin reporting 100\% code coverage for main classes\protect\footnotemark}
		\label{fig:coverage}
	\end{figure}

	To further evaluate the correctness of the algorithm, we have also implemented over 70 test cases for the whole implementation. These were sufficient to achieve 95.2\% overall code coverage. In particular, we paid special attention to the Goal, Pragmatic Goal and Refinement classes as well as to the \texttt{isAchievable} method which were extensively tested until achieving 100\% code coverage and is presented in Figure \ref{fig:coverage}\footnotetext{\url{http://www.eclemma.org/}. Accessed on January 22th, 2015}.

	\subsection{Goal 2: Evaluate the algorithm's runtime usage capability}
	
	\begin{figure*}[htbp]
		\begin{centering}
			\hspace{1cm}
			\subfloat[Average case][Average case\label{fig:scalabilityRandom}]{\includegraphics[width=.38\textwidth]{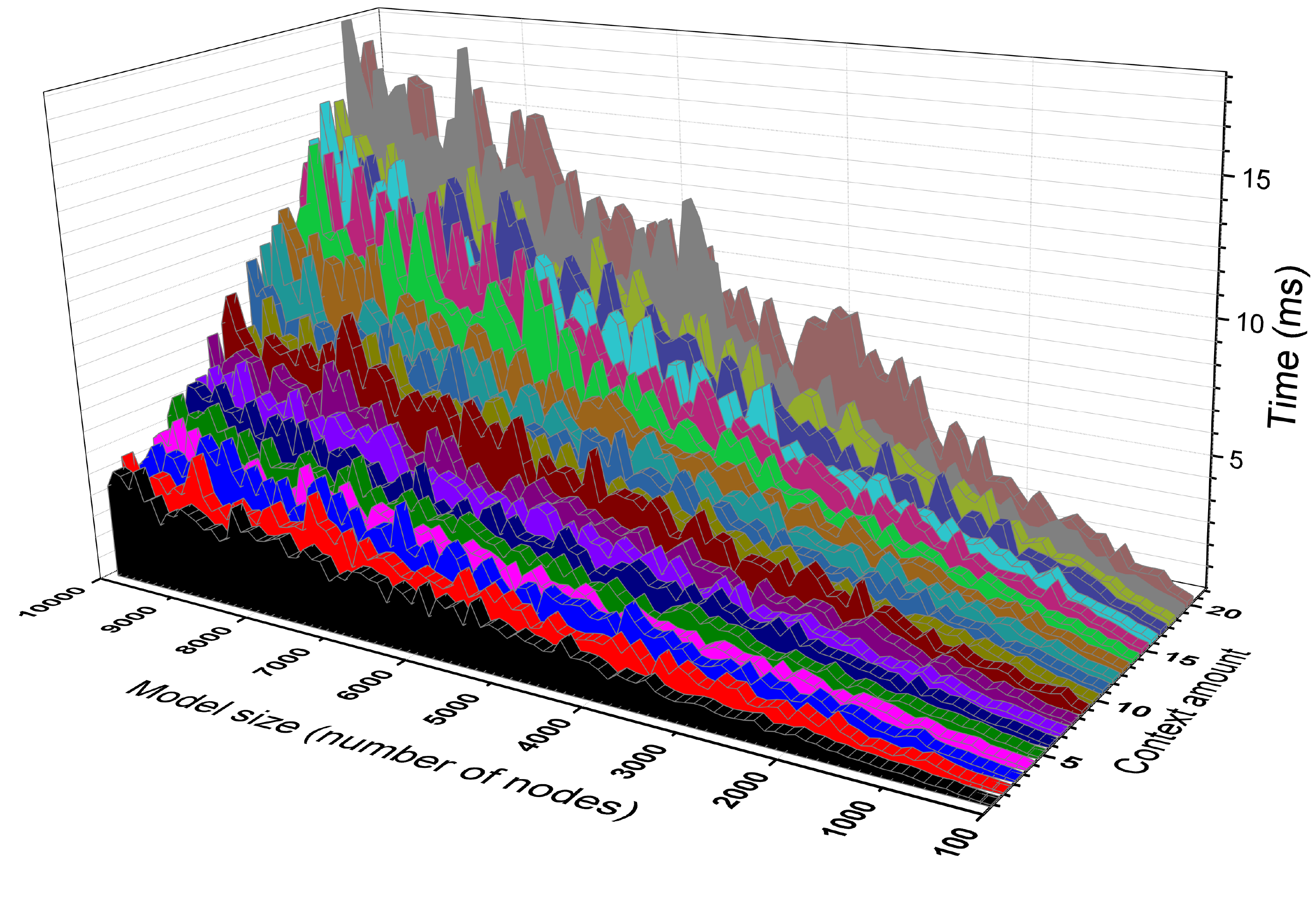}}
			\hfill
			\subfloat[Worst case scenario][Worst case scenario\label{fig:scalabilityWorst}]{\includegraphics[trim=2cm 0.8cm 2cm 3cm,width=.35\textwidth]{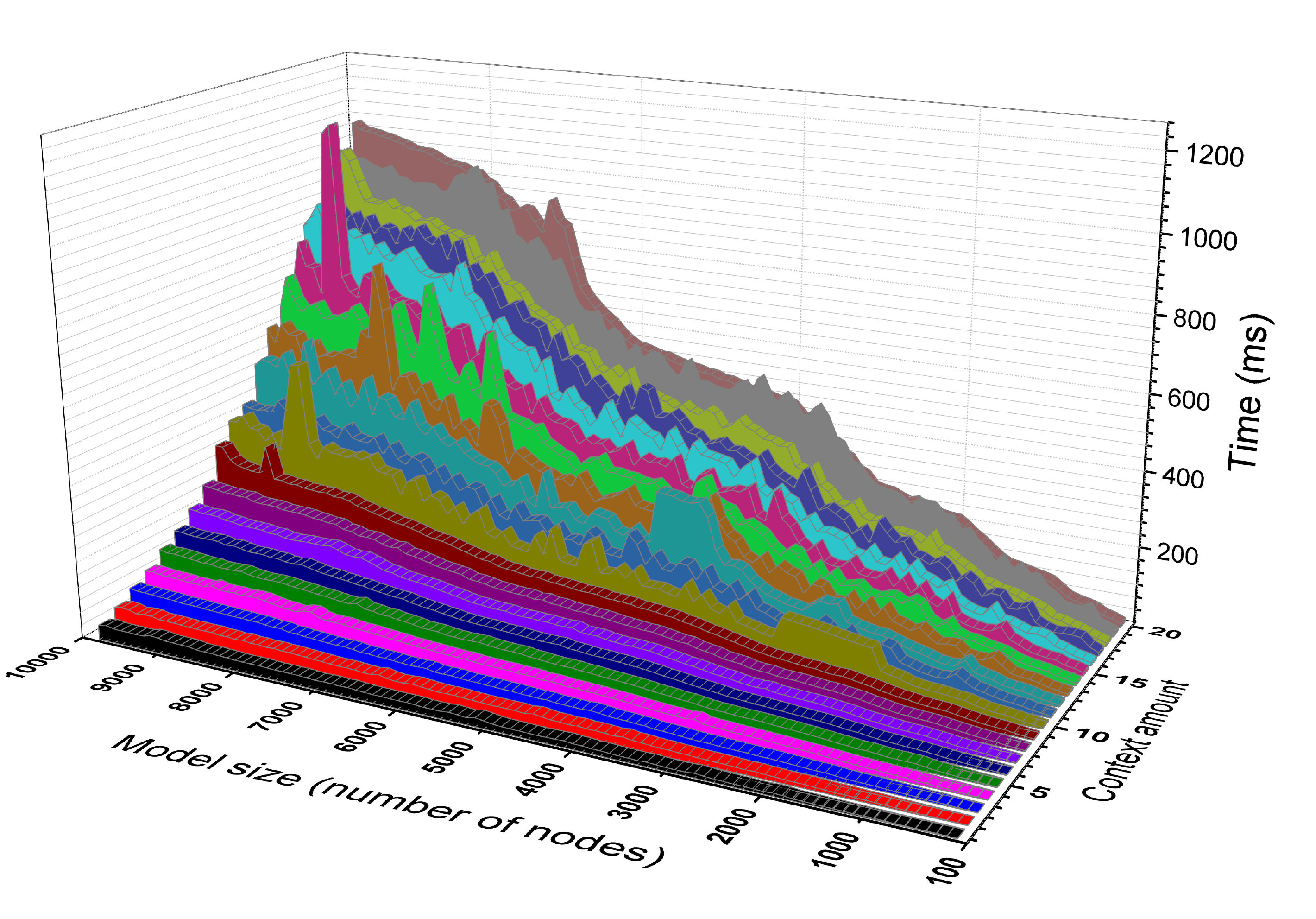}}
			\hspace{1cm}
			\caption{Algorithm's scalability over the model size, in number of nodes, and context amount}
			\label{fig:scalabilityBoth}
		\end{centering}
	\end{figure*}
	
	One of the purposes of this algorithm is to enable the layout, at runtime, of an execution plan which is able to achieve the CGM's root goal. To do so, the algorithm needs to be able to process models with varying complexities, both in terms of context amount and CGM model size, in a reasonable amount of time so that it won't seriously impact the response time. 
	
	\paragraph{Question 2.1 and 2.2: How does the algorithm scale over the amount of goals and contexts in the model in average?}
	
	To evaluate the algorithm's scalability over the model size in terms of goals and contexts amount, we implemented the following test: for each combination of CGM model size (100 to 10000 nodes in steps of 100 nodes) and amount of contexts (1 to 20 contexts), the test would randomly generate 100 CGM models and then the \texttt{isAchievable} method was executed 100 times in each model and the average execution time was measured. Finally, it outputs the average execution time for each combination. The resulting average times are presented in Figure \ref{fig:scalabilityRandom}. 
	
	\paragraph{Question 2.3 and 2.4: How does the algorithm scale over the amount of contexts in the model in average and in the worst case?}
	
	Similarly to the previous experiment, to evaluate the algorithm's scalability in the worst case we implemented another test. This time the generated model would be the algorithm's worst case scenario: an achievable Pragmatic CGM composed solely of AND-Decompositions. This forces the algorithm to traverse the whole tree. The test generated random Pragmatic CGMs with sizes varying from 100 to 10000 nodes and, for each model, performed 100 executions. The observed average time per execution is shown in Figure \ref{fig:scalabilityWorst}. The results were similar in behavior though higher than those observed in the average case.

	\subsection{Goal 3: Algorithm's capability of pinpointing unachievable context sets}
	
	To answer question 3.1 (How many contexts sets can the algorithm evaluate per second for increasingly large models?) we implemented one last test which would generate increasingly larger models with a fixed set of 15 contexts. For each model size 10 random models were generated. Finally, on each of these models and for each combination of the 15 contexts, we have executed the algorithm and measured the percentage of possible context sets it was able to sweep, either finding a suitable solution or not, within ten seconds.
	
\begin{figure}
		\begin{center}
			\includegraphics[width=.44\textwidth, trim=2cm 2cm 2cm 2cm ]{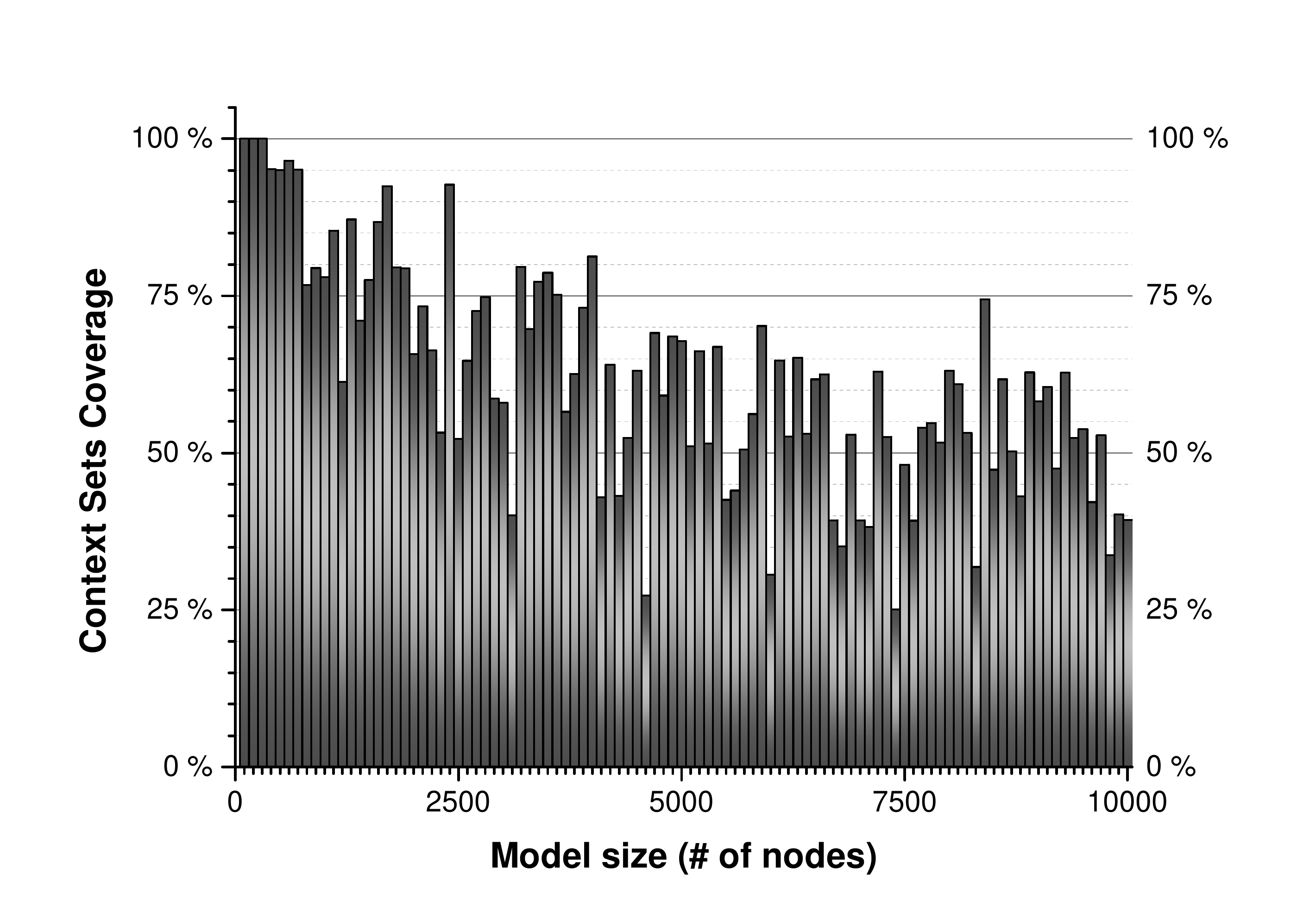}
			\caption{Percentage of possible context sets swept within 10 seconds versus model size} 
			\label{fig:setsCoverage}
		\end{center}
	\end{figure}

	The results can be seen in Figure~\ref{fig:setsCoverage}. As it shows, on smaller models up to 300 goals, the algorithm was able to fully sweep the 32768 context sets within the stipulated deadline. On larger models - up to 5000 goals - the algorithm was able to sweep around 40\% of the combinations. Even for rather large models with 10000 nodes it was able to cover more than 25\% of the possible combinations. 
	
	\subsection{Discussion of the results}
	
	As stated in the GQM plan (Table \ref{tab:GQMPlan}), the experiments had three main goals: (1) determine if there is the necessity for an algorithmic approach, (2) determine if the algorithm could be used at runtime to find a suitable plan for the current context and; (3) whether it could also be used for pinpointing context combinations in which there is no applicable goals/tasks combination sufficiently good to satisfy the pragmatic goals' requirements.
	
	For the first goal, the results have corroborated with the understanding that the Pragmatic CGM complexity is far too great to be dealt simply by the human perspective. 
	On the other hand, the algorithm proved itself as a much faster and deterministic alternative. While the volunteers took up to 17 minutes to provide an answer with 73.19\% reliability, the algorithm produced an answer to the same problem in under 1 millisecond. Also, since it is a deterministic algorithm, the produced answers are always valid. Even if considering very large improvements on the human performance, the algorithmic approach would, most likely, still largely surpass human performance.
	

	Though the comparison between volunteers and an algorithm may seem unfair, the goal of this comparison is to determine the necessity of an algorithmic approach. Being so and as expected, the algorithm has proved itself much better both in terms of efficiency and efficacy as well as subsiding the desired hypothesis that an algorithm is indeed necessary.

	For the second goal, we aimed to evaluate whether the algorithm was suitable for execution at runtime. This meant verifying two main aspects: that the algorithm would efficiently execute even with large Pragmatic CGM models and within a reasonable time to not affect the response time. We performed this analysis for random models (Figures \ref{fig:scalabilityRandom} and \ref{fig:contextSets}) which may have incurred in the observed high variability for the average case. However, an upper boundary was also set with the worst case evaluation (Figure \ref{fig:scalabilityWorst}).
	With regard to the first aspect, Figure \ref{fig:scalabilityBoth} shows that the algorithm's execution time grows linearly over the amount of goals as well as over the amount of contexts in the Pragmatic CGM model, both in the average and in the worst case scenarios.
	The second aspect can also be derived from Figure \ref{fig:scalabilityBoth}. As it shows, even when considering the worst case scenario , the time for evaluating a model with 10000 nodes and 20 contexts was 1081 ms. As a matter of comparison, the default setting of Axis2 (Java's API for web services) for a read timeout is set to be 300 seconds\footnote{\url{http://www-01.ibm.com/support/knowledgecenter/SSD28V_8.5.5/com.ibm.websphere.nd.doc/ae/rwbs_jaxwstimeouts.html}. Accessed on January 14th, 2015}. Therefore, we see this as an acceptable result for the purposes of showing that the algorithm's performance is enough not to severely impact runtime.
	
	Finally, for the third goal we evaluated how long the algorithm took to sweep all context combinations. Given the results from Figure~\ref{fig:setsCoverage}, we can state that for models with up to 10000 goals and 15 contexts, it is actually possible to evaluate all context set combinations within one minute. Thus, the proposed algorithm can indeed be used on a Pragmatic CGM to pinpoint unachievable scenarios with up to 15 contexts, which can then be scrutinized by the CGM designer in order to correct or document it.

	\subsection{Threats to validity}
	Construct validity concerns establishing correct operational measures for the concepts being studied. This was minimized by the usage of the GQM methodology to lay out the evaluation plan. Firstly, the goals that needed to be achieved were laid out; for each goal we envisioned the questions which needed to be answered and only then the metrics were defined with these questions in mind.
	
	Internal validity concerns establishing a causal relationship, whereby certain conditions are shown to lead to other conditions. The experiment setups guaranteed that all computer experiments and evaluations were performed in the same resource and environment. At each evaluation, up to two controlled variables were used.
	
	External validity concerns establishing the domain to which the findings can be generalized. We conducted the assessment in the context of the Mobile Personal Emergency Response, which is specific to a given domain. Nevertheless, future work should assess this in different and real-life domains and with higher number of goals, contexts and quality constraints. In particular, this would help to further assess the scalability we observed. Furthermore, other non-functional properties could be evaluated and we consider it as future work. 
	
	
	\section{Related Work} \label{related}
	
	Previous work have tackled similar problems but to the best of our knowledge none has dealt with the dynamic context-dependent interpretation of requirements and, in particular, of goals. 
	Relevant approaches include the work of 
	Souza and Mylopoulos on Awareness Requirements which are goals that define quality objectives for other goals \cite{Souza2011-1,Souza2011-2};
	the RELAX framework which provides a more rigorous treatment of requirements explicitly related to self-adaptivity, but it does so in a static, yet fuzzy, interpretation \cite{whittle2009relax} ;
	Baresi and Pasquale on Live Goals: goals whose individual behavior changed in order to pursue some qualitative objective and bring the system back to a consistent state \cite{Baresi2010a, Baresi2010} ; 
	Dalpiaz et al. on declarative goals, which are separate goals whose achievement depends on the effects of its refinements on the environment \cite{Dalpiaz2011} and; 
	Sebastiani et al. deal with the Goal-Model satisfiability problem and its mapping into a propositional satisfiability (SAT) problem \cite{sebastiani2004simple}. 
	We argue that the notion of pragmatic goals could enrich the rationale of adaptation proposed in these and other approaches in goal-driven adaptation and that it differs from previous approaches since the above work considered it as a system- or model-wise problem instead of thinking it on a case-to-case context-dependent situation. 
	
	In comparison to the presented work, we differ from Souza and Mylopoulos who consider the quality objective as a distinct goal while we believe it to be an inseparable component of the goal itself: the mere completion of one or all refinements is not enough to achieve a goal, there may be clients' expectations/demands which must be met and which may vary over different contexts. In comparison to the RELAX framework,  we differ in the sense that in our approach a goal's interpretation is not static but context-dependent. We also differ from Baresi and Pasquale in the sense that live goals would react to an inconsistent system state and reason over these particular goals' alternatives while our approach reasons over the whole CGM tree in an effort to, \textit{a priori}, identify and avoid alternatives which will put the system in an inconsistent state, thus maximizing the probability of success. Finally, we differ from Dalpiaz since we believe that the pragmatic nature is not a separate goal but intrinsically related to the goal itself and from Sebastiani since, to enable the algorithm's recursion and obtain a linear complexity algorithm, we use the simplifying assumption that there are no contributions or denials between different goals, thus enabling the treatment of the CGM as a tree and not as a generic graph.

	The novelty of our work in comparison to other approaches in requirements-driven adaptation is twofold:
	(1) The definition of pragmatic goals which means that the satisfaction  criteria for goals is context-dependent.
	(2) The development and implementation of an automated reasoning  that can deterministically answer whether the goal is pragmatically achievable and, if it is, point out an execution plan that is likely to achieve it under the current context.

	\section{Conclusions and Future Work}
	\label{sec:conclusion} 
	
	In this paper we proposed the utilization of a Pragmatic CGM in which the goals' context-dependent interpretation is an integral part of the model. We have also shown why hard goals and soft goals are not enough to grasp some of the real-world peculiarities and context-dependent goal interpretations. 
	
	We defined the pragmatic goals' achievability property. A goal's achievability states whether there is any possible execution plan that fulfills the goal's interpretation under a given context. We also proposed, implemented and evaluated an algorithm to decide on the achievability of a goal and lays out an execution plan. 
	
	We compared the performance of our algorithm to that of a layman's analysis and effectively shown that an algorithmic approach to support the pragmatic goals is needed, considering that human judgment will probably not be fast nor reliable enough. Then, we discussed how the algorithm may enhance requirements engineering by evaluating and pinpointing context sets under which the root goal may not be achieved. 
	
	Finally, we performed a scalability analysis on it and shown that it scales linearly over the amount of goals and context amount. We have also shown that, for models up to 10000 nodes and 20 contexts, our algorithm is able to lay out an execution plan in about a second. Finally, we also evaluated that it was able to sweep all context combinations for models with 15 contexts and up to 10000 nodes in less than a minute, thus making this algorithm also suitable for pinpointing unachievable contexts. 
	
	For future work, we plan to: (1) integrate this algorithm into a CGM modelling tool; (2) study the possibility of the algorithm to return all task sets instead of a single one and (3) how to enhance the model to integrate task dependencies so that it may represent a context-dependent runtime GM with QoS constraints.





\clearpage

\bibliographystyle{IEEEtran}
\bibliography{IEEEabrv,SEAMS}

\begin{thebibliography}{10}
\providecommand{\url}[1]{#1}
\csname url@samestyle\endcsname
\providecommand{\newblock}{\relax}
\providecommand{\bibinfo}[2]{#2}
\providecommand{\BIBentrySTDinterwordspacing}{\spaceskip=0pt\relax}
\providecommand{\BIBentryALTinterwordstretchfactor}{4}
\providecommand{\BIBentryALTinterwordspacing}{\spaceskip=\fontdimen2\font plus
\BIBentryALTinterwordstretchfactor\fontdimen3\font minus
  \fontdimen4\font\relax}
\providecommand{\BIBforeignlanguage}[2]{{%
\expandafter\ifx\csname l@#1\endcsname\relax
\typeout{** WARNING: IEEEtran.bst: No hyphenation pattern has been}%
\typeout{** loaded for the language `#1'. Using the pattern for}%
\typeout{** the default language instead.}%
\else
\language=\csname l@#1\endcsname
\fi
#2}}
\providecommand{\BIBdecl}{\relax}
\BIBdecl

\bibitem{yu1998goals}
E.~Yu and J.~Mylopoulos, ``Why goal-oriented requirements engineering,'' in
  \emph{Proceedings of the 4th International Workshop on Requirements
  Engineering: Foundations of Software Quality}, 1998, pp. 15--22.

\bibitem{Raian2010}
R.~Ali, F.~Dalpiaz, and P.~Giorgini, ``\BIBforeignlanguage{English}{A
  goal-based framework for contextual requirements modeling and analysis},''
  \emph{\BIBforeignlanguage{English}{Requirements Engineering}}, vol.~15,
  no.~4, pp. 439--458, 2010.

\bibitem{tropos}
P.~Bresciani, A.~Perini, P.~Giorgini, F.~Giunchiglia, and J.~Mylopoulos,
  ``Tropos: An agent-oriented software development methodology,''
  \emph{Autonomous Agents and Multi-Agent Systems}, vol.~8, no.~3, pp.
  203--236, 2004.

\bibitem{castro2002towards}
J.~Castro, M.~Kolp, and J.~Mylopoulos, ``Towards requirements-driven
  information systems engineering: the \textit{Tropos} project,''
  \emph{Information systems}, vol.~27, no.~6, pp. 365--389, 2002.

\bibitem{finkelstein2001framework}
A.~Finkelstein and A.~Savigni, ``A framework for requirements engineering for
  context-aware services,'' 2001.

\bibitem{Mendonca2014}
D.~F. Mendon\c{c}a, R.~Ali, and G.~N. Rodrigues, ``Modelling and analysing
  contextual failures for dependability requirements,'' in \emph{Proceedings of
  the 9th International Symposium on Software Engineering for Adaptive and
  Self-Managing Systems}, ser. SEAMS 2014.\hskip 1em plus 0.5em minus
  0.4em\relax New York, NY, USA: ACM, 2014, pp. 55--64.

\bibitem{gqm}
V.~R. Basili, G.~Caldiera, and H.~D. Rombach, ``The goal question metric
  approach,'' in \emph{Encyclopedia of Software Engineering}.\hskip 1em plus
  0.5em minus 0.4em\relax Wiley, 1994.

\bibitem{Souza2011-1}
V.~E. {Silva Souza}, A.~Lapouchnian, W.~N. Robinson, and J.~Mylopoulos,
  ``{Awareness requirements for adaptive systems},'' in \emph{Proceeding of the
  6th International Symposium on Software Engineering for Adaptive and
  Self-Managing Systems - SEAMS 2011}.\hskip 1em plus 0.5em minus 0.4em\relax
  New York, New York, USA: ACM Press, May 2011, p.~60.

\bibitem{Souza2011-2}
V.~E.~S. Souza and J.~Mylopoulos, ``{From awareness requirements to adaptive
  systems: A control-theoretic approach},'' \emph{2011 2nd International
  Workshop on Requirements@Run.Time}, pp. 9--15, Aug. 2011.

\bibitem{whittle2009relax}
J.~Whittle, P.~Sawyer, N.~Bencomo, B.~H. Cheng, and J.-M. Bruel, ``Relax:
  Incorporating uncertainty into the specification of self-adaptive systems,''
  in \emph{Requirements Engineering Conference, 2009. RE'09. 17th IEEE
  International}.\hskip 1em plus 0.5em minus 0.4em\relax IEEE, 2009, pp.
  79--88.

\bibitem{Baresi2010a}
L.~Baresi and L.~Pasquale, ``{Adaptive Goals for Self-Adaptive Service
  Compositions},'' \emph{2010 IEEE International Conference on Web Services},
  pp. 353--360, Jul. 2010.

\bibitem{Baresi2010}
------, ``{Live goals for adaptive service compositions},'' in
  \emph{Proceedings of the 2010 ICSE Workshop on Software Engineering for
  Adaptive and Self-Managing Systems - SEAMS '10}, 2010, pp. 114--123.

\bibitem{Dalpiaz2011}
F.~Dalpiaz, P.~Giorgini, and J.~Mylopoulos, ``{Adaptive socio-technical
  systems: a requirements-based approach},'' \emph{Requirements Engineering},
  vol.~18, no.~1, pp. 1--24, Sep. 2011.

\bibitem{sebastiani2004simple}
R.~Sebastiani, P.~Giorgini, and J.~Mylopoulos, ``Simple and minimum-cost
  satisfiability for goal models,'' in \emph{Advanced Information Systems
  Engineering}.\hskip 1em plus 0.5em minus 0.4em\relax Springer, 2004, pp.
  20--35.

\end{thebibliography}
%
%
%

\end{document}